\documentclass[acmsmall,nonacm,screen]{acmart}  

\usepackage{algorithm}
\usepackage[noend]{algpseudocode}
\usepackage{alltt}
\usepackage{booktabs}
\usepackage[commandnameprefix=ifneeded,final]{changes}  
\usepackage{enumitem}
\usepackage[frozencache,cachedir=.]{minted}
\usepackage{multirow}
\usepackage[autolanguage]{numprint} \npthousandsep{,}
\usepackage{soul}
\usepackage{subcaption}
\usepackage{textcomp}

\graphicspath{{../figures/}{./}}

\makeatletter
\algrenewcommand\ALG@beginalgorithmic{\small}
\algrenewcommand\alglinenumber[1]{\footnotesize #1:}
\makeatother

\makeatletter
\newif\ifanon
\@ifclasswith{acmart}{anonymous}{\anontrue}{\anonfalse}
\makeatother

\newif\ifcomment
\commenttrue    

\newif\iffigabbrv
\figabbrvtrue    
\newcommand{\figtext}{\iffigabbrv Fig.\else Figure\fi}

\newif\ifeqabbrv
\eqabbrvtrue    

\title{Open, Reproducible Calculation of Assembly Indices}

\author{Devansh Vimal}
\orcid{https://orcid.org/0009-0006-2794-8995}
\affiliation{%
    \institution{Arizona State University}
    \department{Biodesign Center for Biocomputing, Security and Society}
    \streetaddress{727 E. Tyler St.}
    \city{Tempe}
    \state{AZ}
    \postcode{85281}
    \country{USA}}
\email{dvpatel5@asu.edu}

\author{Garrett Parzych}
\orcid{https://orcid.org/0009-0008-4789-9603}
\affiliation{%
    \institution{Arizona State University}
    \department{Biodesign Center for Biocomputing, Security and Society}
    \department{School of Computing and Augmented Intelligence}
    \streetaddress{727 E. Tyler St.}
    \city{Tempe}
    \state{AZ}
    \postcode{85281}
    \country{USA}}
\email{gparzych@asu.edu}

\author{Olivia M. Smith}
\orcid{https://orcid.org/0009-0004-2299-3522}
\affiliation{%
    \institution{Arizona State University}
    \department{Biodesign Center for Biocomputing, Security and Society}
    \department{School of Complex Adaptive Systems}
    \streetaddress{727 E. Tyler St.}
    \city{Tempe}
    \state{AZ}
    \postcode{85281}
    \country{USA}}
\email{omsmith8@asu.edu}

\author{Devendra Parkar}
\orcid{https://orcid.org/0009-0009-0133-8875}
\affiliation{%
    \institution{Arizona State University}
    \department{Biodesign Center for Biocomputing, Security and Society}
    \department{School of Computing and Augmented Intelligence}
    \streetaddress{727 E. Tyler St.}
    \city{Tempe}
    \state{AZ}
    \postcode{85281}
    \country{USA}}
\email{dparkar1@asu.edu}

\author{Sean Bergen}
\orcid{https://orcid.org/0009-0004-3570-5120}
\affiliation{%
    \institution{Arizona State University}
    \department{Biodesign Center for Biocomputing, Security and Society}
    \department{School of Computing and Augmented Intelligence}
    \streetaddress{727 E. Tyler St.}
    \city{Tempe}
    \state{AZ}
    \postcode{85281}
    \country{USA}}
\email{sbergen2@asu.edu}

\author{Joshua J. Daymude}
\orcid{https://orcid.org/0000-0001-7294-5626}
\affiliation{%
    \institution{Arizona State University}
    \department{Biodesign Center for Biocomputing, Security and Society}
    \department{School of Computing and Augmented Intelligence}
    \streetaddress{727 E. Tyler St.}
    \city{Tempe}
    \state{AZ}
    \postcode{85281}
    \country{USA}}
\email{jdaymude@asu.edu}

\author{Cole Mathis}
\orcid{https://orcid.org/0000-0001-8424-9169}
\affiliation{%
    \institution{Arizona State University}
    \department{Biodesign Center for Biocomputing, Security and Society}
    \department{School of Complex Adaptive Systems}
    \streetaddress{727 E. Tyler St.}
    \city{Tempe}
    \state{AZ}
    \postcode{85281}
    \country{USA}}
\email{nmathis@asu.edu}

\begin{abstract}
    We present \texttt{assembly-theory}, a Rust package for computing \textit{assembly indices} of covalently bonded molecular structures.
    This is a key complexity measure of \textit{assembly theory}, a recent theoretical framework quantifying selection across diverse systems, most importantly chemistry~\cite{Walker2024-experimentallymeasured,Sharma2023-assemblytheory}.
    \texttt{assembly-theory} is designed for researchers and practitioners alike, providing (i) extensible, high-performance implementations of assembly index calculation algorithms, (ii) comprehensive benchmarks against which current and future algorithmic improvements can be tested, and (iii) Python bindings and \texttt{RDKit}-compatible data loaders to support integration with existing computational pipelines.
\end{abstract}

\begin{document}

\maketitle

\renewcommand{\shortauthors}{Vimal, Parzych, Smith, Parkar, Bergen, Daymude, and Mathis}

\setlength{\parindent}{0pt}
\setlength{\parskip}{0.2\baselineskip}

\section{Background} \label{sec:background}

\begin{figure}[tbh]
    \centering
    \includegraphics[width=\textwidth]{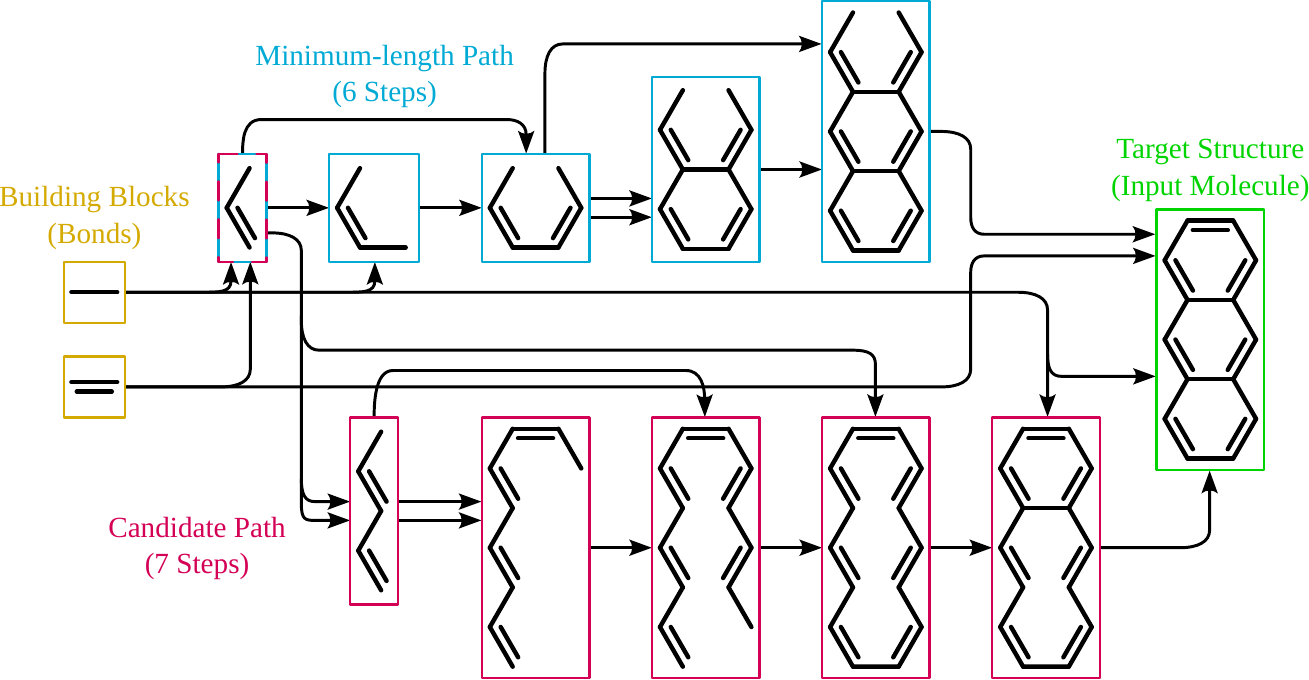}
    \caption{\textit{Assembly Pathways for Anthracene}.
    Starting with bonds as building blocks (yellow), a joining operation yields progressively larger structures by combining any two compatible structures that have already been constructed (arrows).
    These intermediate structures must obey valence rules but otherwise do not have to be physically accessible or chemically synthesizable.
    There may be many assembly pathways from building blocks to a target structure---in this case, Anthracene (green)---but the length of any shortest such pathway (blue) is that structure's assembly index.}
    \label{fig:assemblyindex}
\end{figure}

\textit{Assembly theory} (AT) is a recently developed body of theoretical and empirical work focused on characterizing selection in diverse physical systems, most importantly in chemistry~\cite{Sharma2023-assemblytheory,Walker2024-experimentallymeasured}.
Objects are defined in AT as entities that are finite, distinguishable, decomposable, and persistent in time.
AT characterizes objects based on their \textit{assembly index}, the minimum number of recursive subconstructions required to construct the object starting from a given set of building blocks~\cite{Jirasek2024-investigatingquantifying,Seet2024-rapidcomputation}.
The most commonly studied application domain of AT to date is molecular chemistry, where bonds act as the basic building blocks and the quantity of interest is the \textit{molecular assembly index} (MA); see \figtext~\ref{fig:assemblyindex} for an example.
It has previously been shown that MA can be measured for covalently-bonded molecules using standard analytical techniques such as tandem mass spectrometry as well as infrared and nuclear magnetic resonance spectroscopy~\cite{Jirasek2024-investigatingquantifying}, enabling a novel approach to life detection based on AT~\cite{Marshall2021-identifyingmolecules}.
Beyond life detection, AT and MA have been proposed in methods to generate novel therapeutic drugs, identify environmental pollutants, and gain new insights into evolutionary history by inferring relationships directly from metabolomic data~\cite{Liu2021-exploringmapping,Kahana2024-constructingmolecular}.

\section{Statement of Need} \label{sec:statementofneed}

Despite AT's promising applications, computing MA efficiently remains a challenge.
In general, exact MA calculation is an NP-hard problem~\cite{Kempes2024-assemblytheory}; i.e., the necessary computing resources are likely to grow exponentially with a molecule's number of bonds.
Previous software to compute MA have been approximate, closed-source, platform-dependent, or written in languages rarely used by the broader scientific community.
For example, the original software to compute a split-branch approximation of MA (an upper bound on the exact value) was written in C++ and depended on the MSVC compiler, making it difficult to deploy to non-Windows machines~\cite{Marshall2021-identifyingmolecules}.
Machine learning methods only provide approximate MA values~\cite{Gebhard2022-inferringmolecular}.
The more recent \texttt{assembly\_go} implementation computes MA exactly but is written in Go, yielding worse performance and posing an accessibility barrier for most scientists who are unfamiliar with the language~\cite{Jirasek2024-investigatingquantifying}.
Finally, the latest implementation achieves significant performance milestones through a branch-and-bound approach~\cite{Seet2024-rapidcomputation}.
It is again written in C++ but is not publicly available, prohibiting its use and verification by the community.

With \texttt{assembly-theory}, we provide a high-performance Rust package for MA calculation while also providing Python bindings for key functionality, offering the best efficiency without sacrificing ease of use for scientific practitioners.
We chose Rust for its advantages of cross-platform support, memory-safety, performant runtime, convenient parallelism, and integrated testing and documentation~\cite{Perkel2020-whyscientists}.
By including test and benchmark suites, we also lay a foundation for fair, reproducible comparisons of future algorithmic improvements.

\section{Design and Current Algorithms} \label{sec:design}

\texttt{assembly-theory} is not a single algorithmic implementation of assembly index calculations; rather, it is a framework and source of ground truth within which a diversity of algorithmic approaches can be validated and compared.
We purposely designed \texttt{assembly-theory} with a modular algorithm interface and data structures that can be easily extended to handle new algorithmic developments introduced as AT matures.

Following prior work~\cite{Marshall2021-identifyingmolecules,Jirasek2024-investigatingquantifying,Seet2024-rapidcomputation}, \texttt{assembly-theory} currently implements a top-down approach with two phases that execute in sequence: (1) a search space enumeration and (2) a parallel branch-and-bound search.
The enumeration phase finds all pairs of isomorphic, edge-disjoint subgraphs of the given molecule.
Isomorphic subgraphs are binned into equivalence classes by their \texttt{nauty} canonical representations~\cite{McKay2014-practicalgraph} and pairs of subgraphs within the same class are yielded if they are edge-disjoint.
In the search phase, the given molecule is recursively fragmented by removing duplicate subgraphs enumerated in the first phase.
The MA of smaller fragments are used to determine the MA of their parents.
Optionally, a bounding strategy can be used to improve search efficiency.
We briefly summarize the implemented branch-and-bound ("\texttt{bb}") variants below, but emphasize that \texttt{assembly-theory} is not limited to this top-down, recursive approach.

\begin{itemize}
    \item \texttt{bb-naive} fully enumerates all non-duplicate assembly pathways to find the shortest one.
    \item \texttt{bb-logbound} improves over the naive method by eliminating any assembly pathways whose current length plus $\log_2b$ exceeds the length of the shortest assembly pathway found so far, where $b$ is the number of remaining bonds~\cite{Jirasek2024-investigatingquantifying}.
    \item \texttt{bb-intbound} uses a stronger lower bound on the number of remaining assembly steps provided by an integer addition chain~\cite{Seet2024-rapidcomputation}.
    \item \texttt{bb-allbounds} simultaneously applies the previous integer addition chain bound and a novel bound provided by a vector addition chain.
\end{itemize}

\section{Functionality and Usage Examples} \label{sec:usage}

\texttt{assembly-theory} can be used to compute assembly indices as a standalone executable, as a library imported by other Rust code, or via a Python interface.
Here, we provide usage examples of each; in the next section, we demonstrate testing and benchmarking functionality.

\subsection{Standalone Executable} \label{subsec:standalone}

Rust provides the \texttt{cargo} build system and package manager for dependency management, compilation, packaging, and versioning and the \href{https://crates.io/crates/assembly-theory}{crates.io} registry for package distribution.
To install the standalone executable, run:

\begin{minted}{shell}
    > cargo install assembly-theory
\end{minted}

This executable takes as input a path to a \texttt{.mol} file and returns that molecule's assembly index:

\begin{minted}{shell}
    > assembly-theory data/checks/anthracene.mol
    6
\end{minted}

Running with the \verb|--verbose| flag provides additional information, including the input molecule's \textit{number of disjoint, isomorphic subgraph pairs} (i.e., the number of times any molecular substructure is repeated inside the molecule) and the size of the top-down algorithm's \textit{search space} (i.e., its total number of recursive calls).

\begin{minted}{shell}
    > assembly-theory data/checks/anthracene.mol --verbose
    Assembly Index: 6
    Duplicate subgraph pairs: 406
    Search Space: 2306
\end{minted}

By default, \texttt{assembly-theory} parallelizes its recursive search over as many threads as the OS allows.
To disable parallelism, use the \verb|--serial| flag.

Also by default, \texttt{assembly-theory} uses its fastest algorithm for assembly index calculation (currently \texttt{bb-allbounds}, see the previous section).
To use a specific bound or disable bounds altogether, set the \verb|--bounds| or \verb|--no-bounds| flags:

\begin{minted}{shell}
    # bb-naive, no bounds
    > assembly-theory data/checks/anthracene.mol --no-bounds

    # bb-logbound, only logarithmic bound (Jirasek et al., 2024)
    > assembly-theory data/checks/anthracene.mol --bounds log

    # bb-intbound, only integer addition chain bound (Seet et al., 2024)
    > assembly-theory data/checks/anthracene.mol --bounds int-chain

    # bb-allbounds, both integer and vector addition chain bounds
    > assembly-theory data/checks/anthracene.mol --bounds int-chain vec-chain
\end{minted}

Finally, the \verb|--molecule-info| flag prints the molecule's graph representation as a vertex and edge list, the \verb|--help| flag prints a guide to this command line interface, and the \verb|--version| flag prints the current \texttt{assembly-theory} version.

\subsection{Rust Library} \label{subsec:rustlibrary}

To include \texttt{assembly-theory} in a broader Rust project, run:

\begin{minted}{shell}
    > cargo add assembly-theory
\end{minted}

Complete documentation of the library is available on \href{https://docs.rs/assembly-theory/latest/assembly_theory}{docs.rs}; a simple usage example is:

\begin{minted}{rust}
    use assembly_theory::*;

    // Read a .mol file as an assembly_theory::molecule::Molecule.
    let molfile = fs::read_to_string(path)?;
    let anthracene = loader::parse_molfile_str(&molfile).expect("Parsing failed");

    // Calculate the molecule's assembly index.
    assert_eq!(assembly::index(&anthracene), 6);
\end{minted}

\subsection{Python Interface} \label{subsec:python}

We use \href{https://github.com/PyO3/maturin}{\texttt{maturin}} to repackage the \texttt{assembly-theory} Rust binaries as the \texttt{assembly\_theory} package for Python.
Instructions for this build process can be found in our \texttt{README}; otherwise, the Python package can be obtained from PyPI in the usual way:

\begin{minted}{shell}
    > pip install assembly_theory
\end{minted}

Our Python interface is built for compatibility with \texttt{RDKit}, the standard Python library for cheminformatics~\cite{2024-rdkitopensource}.
Molecules can be loaded and manipulated using the \texttt{rdkit.Chem.Mol} class and then passed to our functions for assembly index calculation:

\begin{minted}{python}
    >>> import assembly_theory as at
    >>> from rdkit import Chem
    >>> anthracene = Chem.MolFromSmiles("c1ccc2cc3ccccc3cc2c1")
    >>> at.molecular_assembly(anthracene)
    6
\end{minted}

In detail, \texttt{assembly\_theory} exposes three main functions:

\begin{enumerate}
    \item \textbf{\texttt{molecular\_assembly}}\texttt{(mol: Chem.Mol, bounds: set[str] = None, no\_bounds: bool = False, timeout: int = None, serial: bool = False) -> int}
    \item \textbf{\texttt{molecular\_assembly\_verbose}}\texttt{(mol: Chem.Mol, bounds: set[str] = None, no\_bounds: bool = False, timeout: int = None, serial: bool = False) -> dict}
    \item \textbf{\texttt{molecule\_info}}\texttt{(mol: Chem.Mol) -> str}
\end{enumerate}

These correspond to (1) running the Rust \texttt{assembly-theory} executable to obtain only an assembly index, (2) running with the \verb|--verbose| flag to also obtain the number of disjoint isomorphic subgraph pairs (\texttt{duplicates}) and search space size (\texttt{space}), and (3) running with the \verb|--molecule-info| flag to obtain molecule information, respectively.
The \texttt{timeout} parameter is specific to the Python interface: when set to a non-\texttt{None} integer value, a \texttt{TimeoutError} is raised if assembly index calculation exceeds \texttt{timeout} seconds.

\section{Tests and Benchmarks} \label{sec:testbench}

\texttt{assembly-theory} includes test and benchmark suites for software validation and performance evaluation, respectively.
Both use curated reference datasets representing different classes of molecules, arranged roughly in order of increasing molecular size and complexity:

\begin{itemize}
    \item \texttt{gdb13\_1201}: \numprint{1201} small, organic molecular structures sampled from GDB-13, a database of enumerated chemical structures containing Carbon, Hydrogen, Nitrogen, Oxygen, Sulfur, and Chlorine that are constrained only by valence rules and quantum mechanics but may not be chemically stable or synthesizable~\cite{Blum2009-970million}.
    Our sample includes all 201 molecules with 4--5 heavy atoms and 200 randomly sampled molecules for each number of heavy atoms from 6--10.
    These molecules' MA range from 2--9.

    \item \texttt{gdb17\_200}: 200 organic molecular structures sampled from the GDB-17 database, which includes additional nuclei beyond GDB-13 such as the halogens Flourine and Iodine~\cite{Ruddigkeit2012-enumeration166}.
    Compared to GDB-13, these molecules are typically larger and represent more structural diversity.
    Our sample includes 50 randomly sampled molecules for each number of heavy atoms from 14--17.
    These molecules' MA range from 7--16.

    \item \texttt{checks}: 15 named molecules (e.g., anthracene, aspirin, caffeine, morphine) from KEGG COMPOUND~\cite{Kanehisa2000-keggkyoto,Kanehisa2019-understandingorigin,Kanehisa2023-keggtaxonomybased} primarily used for rapid testing.
    These molecules' number of heavy atoms range from 5--21 and have MA from 3--14.

    \item \texttt{coconut\_55}: 55 natural products sampled from the COCONUT database~\cite{Sorokina2021-coconutonline}, accessed in late 2024, prior to COCONUT 2.0~\cite{Chandrasekhar2025-coconut20}.
    Natural products (or secondary metabolites) are a rich source of evolved chemical complexity, often exhibiting drug-like properties.
    Subsets of this database were used to benchmark recent algorithmic progress in~\cite{Seet2024-rapidcomputation}.
    Our sample includes five randomly sampled molecules for each number of heavy atoms from 15--25.
    These molecules' MA range from 7--16.
\end{itemize}

We curated these reference datasets for their structural diversity and approachable runtime on commodity hardware.
Larger, more demanding datasets can be added as needed.

The \texttt{assembly-theory} test suite contains unit tests validating internal functionality and integration tests verifying the calculation of correct assembly indices for all molecules in our reference datasets.
Each reference dataset contains an \texttt{ma-index.csv} file with ground truth assembly indices calculated using the closed-source~\cite{Seet2024-rapidcomputation} algorithm, privately provided to us by the authors for this use only.

\begin{table}[tbh]
    \centering
    \caption{Mean benchmark execution times for \texttt{assembly\_go}~\cite{Jirasek2024-investigatingquantifying} vs.\ \texttt{assembly-theory} across reference datasets.
    The benchmark times the MA calculation of all molecules in a given dataset in sequence, excluding the time required to parse and load \texttt{.mol} files into internal molecular graph representations.
    \texttt{assembly\_go} uses its default parameters while \texttt{assembly-theory} tests each of its algorithm variants independently.
    Each benchmark was run on a Linux machine with a 5.7 GHz Ryzen 9 7950X CPU (16 cores) and 64 GB of memory.
    Means are reported over 20 samples per software--dataset pair, except those marked with an $\ast$ which have prohibitively long runtimes and thus ran only once.}
    \label{tab:benchtimes}
    \begin{tabular}{lrrrrr}
        \toprule
         & \texttt{assembly\_go} & \texttt{bb-naive} & \texttt{bb-logbound} & \texttt{bb-intbound} & \texttt{bb-allbounds} \\
        \midrule
        \texttt{gdb13\_1201} & 0.968 s & 0.147 s & 0.149 s & \textbf{0.140 s} & 0.142 s \\
        \texttt{gdb17\_200} & 46.189 s & 4.586 s & 3.446 s & 2.988 s & \textbf{2.946 s} \\
        \texttt{checks} & 212.000 s & 12.635 s & 1.767 s & 1.357 s & \textbf{1.297 s} \\
        \texttt{coconut\_55} & 1.48 h$^\ast$ & 175.687 s & 57.123 s & 13.188 s & \textbf{12.996 s} \\
        \bottomrule
    \end{tabular}
\end{table}

Our benchmark suite evaluates \texttt{assembly-theory} performance by running repeated assembly index calculations over individual molecules or entire reference datasets.
We leverage the \href{https://bheisler.github.io/criterion.rs/criterion/}{\texttt{criterion}} package for Rust to automatically collect detailed timing statistics, charts, and estimates of performance improvements and regressions.
As an example, Table~\ref{tab:benchtimes} shows \texttt{assembly-theory} performance across our four reference datasets against that of \texttt{assembly\_go}~\cite{Jirasek2024-investigatingquantifying}.
Depending on the dataset and choice of \texttt{assembly-theory} algorithm, \texttt{assembly-theory} outperforms \texttt{assembly\_go} by one or two orders of magnitude.
The 6.5--120.0x speedup of \texttt{bb-logbound} over \texttt{assembly\_go} most clearly represents the efficiency of Rust over Go, since both use the same branch-and-bound approach with a logarithmic bound.
Algorithmic improvements such as the \texttt{bb-allbounds} combination of an integer addition chain bound~\cite{Seet2024-rapidcomputation} and our novel vector addition chain bound yield more dramatic speedups for larger molecules, like those up to 410.0x for \texttt{coconut\_55}.
This internal comparison showcases \texttt{assembly-theory} as a framework capable of comparing multiple algorithmic approaches on equal footing, free of differences in underlying datasets or language-specific efficiency issues.

\begin{figure}[tbh]
    \centering
    \includegraphics[width=0.75\textwidth]{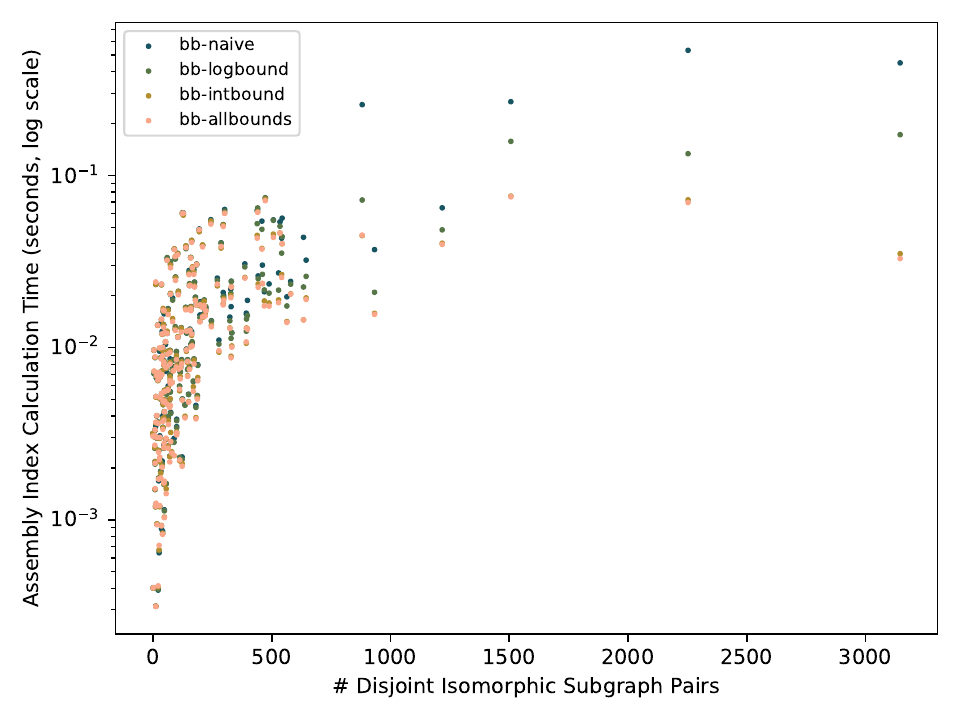}
    \caption{\textit{Per-Molecule Benchmark Times}.
    The mean assembly index calculation time across 20 samples for each molecule (dot) in \texttt{gdb17\_200} as a function of the molecule's number of duplicate isomorphic subgraphs, a measure roughly correlated with the molecule's size and complexity.
    The same four \texttt{assembly-theory} branch-and-bound algorithms from \figtext~\ref{tab:benchtimes} are shown here.}
    \label{fig:timescatter}
\end{figure}

If finer-grained timing insights are needed, \texttt{assembly-theory} can also benchmark assembly index calculations for each individual molecule in a reference dataset.
For example, \figtext~\ref{fig:timescatter} shows the calculation time of each molecule in \texttt{gdb17\_200} for the four branch-and-bound algorithms.
This is useful for teasing out which molecules are "hard" and characterizing where algorithmic improvements make the largest impact.

\section{Availability and Governance} \label{sec:governance}

\texttt{assembly-theory} source code and documentation are openly available on \href{https://github.com/DaymudeLab/assembly-theory}{GitHub}.
Following the standard practice for Rust packages, \texttt{assembly-theory} is dual-licensed under the MIT and Apache-2.0 licenses.
External feedback and code contributions are handled through the usual Issues and Pull Request interfaces; guidelines for contributions are listed in \texttt{HACKING.md}.
The project's \textit{maintainers} (initially Vimal, Daymude, and Mathis) will govern the project using the committee model: high-level decisions about the project's direction require maintainer consensus, major code changes require majority approval, hotfixes and patches require at least one approval, new maintainers may be added by unanimous decision of the existing maintainers, and existing maintainers may step down with advance notice.

\section*{Author Contributions}

GP, DV, and CM formalized the branch-and-bound algorithm design.
GP and SB formalized the integer and vector addition chain bounds.
DV was the primary software developer (architecture, command line interface, molecule representations, unit tests, parallelism, performance engineering).
GP implemented all bound calculations.
DP and DV implemented the \texttt{.mol} file parser and dataset-based benchmarks.
CM implemented the Python interface.
OMS curated all reference datasets and assembly index ground truths with input from CM.
SB and JJD wrote the \texttt{assembly\_go} benchmarks.
JJD conducted and analyzed the benchmarks shown in Table~\ref{tab:benchtimes}.
DP, SB, GP, and JJD produced \figtext~\ref{fig:timescatter}.
JJD and CM wrote the paper.

\begin{acks}
    GP and JJD are supported in part by NSF award CCF-2312537.
    DV, OMS, and CM acknowledge support from the ASU Biodesign Institute.
\end{acks}

\bibliographystyle{ACM-Reference-Format}
\bibliography{ref}


\begin{thebibliography}{19}


\ifx \showCODEN    \undefined \def \showCODEN     #1{\unskip}     \fi
\ifx \showDOI      \undefined \def \showDOI       #1{#1}\fi
\ifx \showISBNx    \undefined \def \showISBNx     #1{\unskip}     \fi
\ifx \showISBNxiii \undefined \def \showISBNxiii  #1{\unskip}     \fi
\ifx \showISSN     \undefined \def \showISSN      #1{\unskip}     \fi
\ifx \showLCCN     \undefined \def \showLCCN      #1{\unskip}     \fi
\ifx \shownote     \undefined \def \shownote      #1{#1}          \fi
\ifx \showarticletitle \undefined \def \showarticletitle #1{#1}   \fi
\ifx \showURL      \undefined \def \showURL       {\relax}        \fi
\providecommand\bibfield[2]{#2}
\providecommand\bibinfo[2]{#2}
\providecommand\natexlab[1]{#1}
\providecommand\showeprint[2][]{arXiv:#2}

\bibitem[202(2024)]%
        {2024-rdkitopensource}
 \bibinfo{year}{2024}\natexlab{}.
\newblock \bibinfo{title}{{{RDKit}}: Open-Source Cheminformatics}.
\newblock
\newblock
\urldef\tempurl%
\url{https://doi.org/10.5281/zenodo.591637}
\showDOI{\tempurl}


\bibitem[Blum and Reymond(2009)]%
        {Blum2009-970million}
\bibfield{author}{\bibinfo{person}{Lorenz~C. Blum} {and}
  \bibinfo{person}{Jean-Louis Reymond}.} \bibinfo{year}{2009}\natexlab{}.
\newblock \showarticletitle{970 {{Million Druglike Small Molecules}} for
  {{Virtual Screening}} in the {{Chemical Universe Database GDB-13}}}.
\newblock \bibinfo{journal}{\emph{Journal of the American Chemical Society}}
  \bibinfo{volume}{131}, \bibinfo{number}{25} (\bibinfo{year}{2009}),
  \bibinfo{pages}{8732--8733}.
\newblock
\urldef\tempurl%
\url{https://doi.org/10.1021/ja902302h}
\showDOI{\tempurl}


\bibitem[Chandrasekhar et~al\mbox{.}(2025)]%
        {Chandrasekhar2025-coconut20}
\bibfield{author}{\bibinfo{person}{Venkata Chandrasekhar},
  \bibinfo{person}{Kohulan Rajan}, \bibinfo{person}{Sri~Ram~Sagar Kanakam},
  \bibinfo{person}{Nisha Sharma}, \bibinfo{person}{Viktor Wei{\ss}enborn},
  \bibinfo{person}{Jonas Schaub}, {and} \bibinfo{person}{Christoph Steinbeck}.}
  \bibinfo{year}{2025}\natexlab{}.
\newblock \showarticletitle{{{COCONUT}} 2.0: A Comprehensive Overhaul and
  Curation of the Collection of Open Natural Products Database}.
\newblock \bibinfo{journal}{\emph{Nucleic Acids Research}}
  \bibinfo{volume}{53}, \bibinfo{number}{D1} (\bibinfo{year}{2025}),
  \bibinfo{pages}{D634--D643}.
\newblock
\urldef\tempurl%
\url{https://doi.org/10.1093/nar/gkae1063}
\showDOI{\tempurl}


\bibitem[Gebhard et~al\mbox{.}(2022)]%
        {Gebhard2022-inferringmolecular}
\bibfield{author}{\bibinfo{person}{Timothy~D. Gebhard},
  \bibinfo{person}{Aaron~C. Bell}, \bibinfo{person}{Jian Gong},
  \bibinfo{person}{Jaden J.~A. Hastings}, \bibinfo{person}{G.~Matthew Fricke},
  \bibinfo{person}{Nathalie Cabrol}, \bibinfo{person}{Scott Sandford},
  \bibinfo{person}{Michael Phillips}, \bibinfo{person}{Kimberley
  {Warren-Rhodes}}, {and} \bibinfo{person}{At{\i}l{\i}m~G{\"u}ne{\c s}
  Baydin}.} \bibinfo{year}{2022}\natexlab{}.
\newblock \showarticletitle{Inferring Molecular Complexity from Mass
  Spectrometry Data Using Machine Learning}. In
  \bibinfo{booktitle}{\emph{Machine {{Learning}} and the {{Physical Sciences
  Workshop}} at {{NeurIPS}} 2022}}. \bibinfo{pages}{1--7}.
\newblock


\bibitem[Jirasek et~al\mbox{.}(2024)]%
        {Jirasek2024-investigatingquantifying}
\bibfield{author}{\bibinfo{person}{Michael Jirasek}, \bibinfo{person}{Abhishek
  Sharma}, \bibinfo{person}{Jessica~R. Bame}, \bibinfo{person}{S.~Hessam~M.
  Mehr}, \bibinfo{person}{Nicola Bell}, \bibinfo{person}{Stuart~M. Marshall},
  \bibinfo{person}{Cole Mathis}, \bibinfo{person}{Alasdair MacLeod},
  \bibinfo{person}{Geoffrey J.~T. Cooper}, \bibinfo{person}{Marcel Swart},
  \bibinfo{person}{Rosa Mollfulleda}, {and} \bibinfo{person}{Leroy Cronin}.}
  \bibinfo{year}{2024}\natexlab{}.
\newblock \showarticletitle{Investigating and {{Quantifying Molecular
  Complexity Using Assembly Theory}} and {{Spectroscopy}}}.
\newblock \bibinfo{journal}{\emph{ACS Central Science}} \bibinfo{volume}{10},
  \bibinfo{number}{5} (\bibinfo{year}{2024}), \bibinfo{pages}{1054--1064}.
\newblock
\urldef\tempurl%
\url{https://doi.org/10.1021/acscentsci.4c00120}
\showDOI{\tempurl}


\bibitem[Kahana et~al\mbox{.}(2024)]%
        {Kahana2024-constructingmolecular}
\bibfield{author}{\bibinfo{person}{Amit Kahana}, \bibinfo{person}{Alasdair
  MacLeod}, \bibinfo{person}{Hessam Mehr}, \bibinfo{person}{Abhishek Sharma},
  \bibinfo{person}{Emma Carrick}, \bibinfo{person}{Michael Jirasek},
  \bibinfo{person}{Sara~I. Walker}, {and} \bibinfo{person}{Leroy Cronin}.}
  \bibinfo{year}{2024}\natexlab{}.
\newblock \bibinfo{title}{Constructing the {{Molecular Tree}} of {{Life}} Using
  {{Assembly Theory}} and {{Mass Spectrometry}}}.
\newblock
\newblock
\urldef\tempurl%
\url{https://doi.org/10.48550/arxiv.2408.09305}
\showDOI{\tempurl}
\showeprint[arxiv]{2408.09305}


\bibitem[Kanehisa(2019)]%
        {Kanehisa2019-understandingorigin}
\bibfield{author}{\bibinfo{person}{Minoru Kanehisa}.}
  \bibinfo{year}{2019}\natexlab{}.
\newblock \showarticletitle{Toward Understanding the Origin and Evolution of
  Cellular Organisms}.
\newblock \bibinfo{journal}{\emph{Protein Science}} \bibinfo{volume}{28},
  \bibinfo{number}{11} (\bibinfo{year}{2019}), \bibinfo{pages}{1947--1951}.
\newblock
\urldef\tempurl%
\url{https://doi.org/10.1002/pro.3715}
\showDOI{\tempurl}


\bibitem[Kanehisa et~al\mbox{.}(2023)]%
        {Kanehisa2023-keggtaxonomybased}
\bibfield{author}{\bibinfo{person}{Minoru Kanehisa}, \bibinfo{person}{Miho
  Furumichi}, \bibinfo{person}{Yoko Sato}, \bibinfo{person}{Masayuki
  Kawashima}, {and} \bibinfo{person}{Mari {Ishiguro-Watanabe}}.}
  \bibinfo{year}{2023}\natexlab{}.
\newblock \showarticletitle{{{KEGG}} for Taxonomy-Based Analysis of Pathways
  and Genomes}.
\newblock \bibinfo{journal}{\emph{Nucleic Acids Research}}
  \bibinfo{volume}{51}, \bibinfo{number}{D1} (\bibinfo{year}{2023}),
  \bibinfo{pages}{D587--D592}.
\newblock
\urldef\tempurl%
\url{https://doi.org/10.1093/nar/gkac963}
\showDOI{\tempurl}


\bibitem[Kanehisa and Goto(2000)]%
        {Kanehisa2000-keggkyoto}
\bibfield{author}{\bibinfo{person}{Minoru Kanehisa} {and}
  \bibinfo{person}{Susumu Goto}.} \bibinfo{year}{2000}\natexlab{}.
\newblock \showarticletitle{{{KEGG}}: {{Kyoto Encyclopedia}} of {{Genes}} and
  {{Genomes}}}.
\newblock \bibinfo{journal}{\emph{Nucleic Acids Research}}
  \bibinfo{volume}{28}, \bibinfo{number}{1} (\bibinfo{year}{2000}),
  \bibinfo{pages}{27--30}.
\newblock
\urldef\tempurl%
\url{https://doi.org/10.1093/nar/28.1.27}
\showDOI{\tempurl}


\bibitem[Kempes et~al\mbox{.}(2024)]%
        {Kempes2024-assemblytheory}
\bibfield{author}{\bibinfo{person}{Christopher Kempes},
  \bibinfo{person}{Sara~I. Walker}, \bibinfo{person}{Michael Lachmann}, {and}
  \bibinfo{person}{Leroy Cronin}.} \bibinfo{year}{2024}\natexlab{}.
\newblock \bibinfo{title}{Assembly {{Theory}} and Its {{Relationship}} with
  {{Computational Complexity}}}.
\newblock
\newblock
\urldef\tempurl%
\url{https://doi.org/10.48550/arXiv.2406.12176}
\showDOI{\tempurl}
\showeprint[arxiv]{2406.12176}


\bibitem[Liu et~al\mbox{.}(2021)]%
        {Liu2021-exploringmapping}
\bibfield{author}{\bibinfo{person}{Yu Liu}, \bibinfo{person}{Cole Mathis},
  \bibinfo{person}{Micha{\l}~Dariusz Bajczyk}, \bibinfo{person}{Stuart~M.
  Marshall}, \bibinfo{person}{Liam Wilbraham}, {and} \bibinfo{person}{Leroy
  Cronin}.} \bibinfo{year}{2021}\natexlab{}.
\newblock \showarticletitle{Exploring and Mapping Chemical Space with Molecular
  Assembly Trees}.
\newblock \bibinfo{journal}{\emph{Science Advances}} \bibinfo{volume}{7},
  \bibinfo{number}{39} (\bibinfo{year}{2021}), \bibinfo{pages}{eabj2465}.
\newblock
\urldef\tempurl%
\url{https://doi.org/10.1126/sciadv.abj2465}
\showDOI{\tempurl}


\bibitem[Marshall et~al\mbox{.}(2021)]%
        {Marshall2021-identifyingmolecules}
\bibfield{author}{\bibinfo{person}{Stuart~M. Marshall}, \bibinfo{person}{Cole
  Mathis}, \bibinfo{person}{Emma Carrick}, \bibinfo{person}{Graham Keenan},
  \bibinfo{person}{Geoffrey J.~T. Cooper}, \bibinfo{person}{Heather Graham},
  \bibinfo{person}{Matthew Craven}, \bibinfo{person}{Piotr~S. Gromski},
  \bibinfo{person}{Douglas~G. Moore}, \bibinfo{person}{{\relax Sara}.~I.
  Walker}, {and} \bibinfo{person}{Leroy Cronin}.}
  \bibinfo{year}{2021}\natexlab{}.
\newblock \showarticletitle{Identifying Molecules as Biosignatures with
  Assembly Theory and Mass Spectrometry}.
\newblock \bibinfo{journal}{\emph{Nature Communications}} \bibinfo{volume}{12},
  \bibinfo{number}{1} (\bibinfo{year}{2021}), \bibinfo{pages}{3033}.
\newblock
\urldef\tempurl%
\url{https://doi.org/10.1038/s41467-021-23258-x}
\showDOI{\tempurl}


\bibitem[McKay and Piperno(2014)]%
        {McKay2014-practicalgraph}
\bibfield{author}{\bibinfo{person}{Brendan~D. McKay} {and}
  \bibinfo{person}{Adolfo Piperno}.} \bibinfo{year}{2014}\natexlab{}.
\newblock \showarticletitle{Practical Graph Isomorphism, {{II}}}.
\newblock \bibinfo{journal}{\emph{Journal of Symbolic Computation}}
  \bibinfo{volume}{60} (\bibinfo{year}{2014}), \bibinfo{pages}{94--112}.
\newblock
\urldef\tempurl%
\url{https://doi.org/10.1016/j.jsc.2013.09.003}
\showDOI{\tempurl}


\bibitem[Perkel(2020)]%
        {Perkel2020-whyscientists}
\bibfield{author}{\bibinfo{person}{Jeffrey~M. Perkel}.}
  \bibinfo{year}{2020}\natexlab{}.
\newblock \showarticletitle{Why Scientists Are Turning to {{Rust}}}.
\newblock \bibinfo{journal}{\emph{Nature}} \bibinfo{volume}{588},
  \bibinfo{number}{7836} (\bibinfo{year}{2020}), \bibinfo{pages}{185--186}.
\newblock
\urldef\tempurl%
\url{https://doi.org/10.1038/d41586-020-03382-2}
\showDOI{\tempurl}


\bibitem[Ruddigkeit et~al\mbox{.}(2012)]%
        {Ruddigkeit2012-enumeration166}
\bibfield{author}{\bibinfo{person}{Lars Ruddigkeit}, \bibinfo{person}{Ruud
  Van~Deursen}, \bibinfo{person}{Lorenz~C. Blum}, {and}
  \bibinfo{person}{Jean-Louis Reymond}.} \bibinfo{year}{2012}\natexlab{}.
\newblock \showarticletitle{Enumeration of 166 {{Billion Organic Small
  Molecules}} in the {{Chemical Universe Database GDB-17}}}.
\newblock \bibinfo{journal}{\emph{Journal of Chemical Information and
  Modeling}} \bibinfo{volume}{52}, \bibinfo{number}{11} (\bibinfo{year}{2012}),
  \bibinfo{pages}{2864--2875}.
\newblock
\urldef\tempurl%
\url{https://doi.org/10.1021/ci300415d}
\showDOI{\tempurl}


\bibitem[Seet et~al\mbox{.}(2024)]%
        {Seet2024-rapidcomputation}
\bibfield{author}{\bibinfo{person}{Ian Seet}, \bibinfo{person}{Keith~Y.
  Patarroyo}, \bibinfo{person}{Gage Siebert}, \bibinfo{person}{Sara~I. Walker},
  {and} \bibinfo{person}{Leroy Cronin}.} \bibinfo{year}{2024}\natexlab{}.
\newblock \bibinfo{title}{Rapid {{Computation}} of the {{Assembly Index}} of
  {{Molecular Graphs}}}.
\newblock
\newblock
\urldef\tempurl%
\url{https://doi.org/10.48550/arXiv.2410.09100}
\showDOI{\tempurl}
\showeprint[arxiv]{2410.09100}


\bibitem[Sharma et~al\mbox{.}(2023)]%
        {Sharma2023-assemblytheory}
\bibfield{author}{\bibinfo{person}{Abhishek Sharma},
  \bibinfo{person}{D{\'a}niel Cz{\'e}gel}, \bibinfo{person}{Michael Lachmann},
  \bibinfo{person}{Christopher~P. Kempes}, \bibinfo{person}{Sara~I. Walker},
  {and} \bibinfo{person}{Leroy Cronin}.} \bibinfo{year}{2023}\natexlab{}.
\newblock \showarticletitle{Assembly Theory Explains and Quantifies Selection
  and Evolution}.
\newblock \bibinfo{journal}{\emph{Nature}} \bibinfo{volume}{622},
  \bibinfo{number}{7982} (\bibinfo{year}{2023}), \bibinfo{pages}{321--328}.
\newblock
\urldef\tempurl%
\url{https://doi.org/10.1038/s41586-023-06600-9}
\showDOI{\tempurl}


\bibitem[Sorokina et~al\mbox{.}(2021)]%
        {Sorokina2021-coconutonline}
\bibfield{author}{\bibinfo{person}{Maria Sorokina}, \bibinfo{person}{Peter
  Merseburger}, \bibinfo{person}{Kohulan Rajan}, \bibinfo{person}{Mehmet~Aziz
  Yirik}, {and} \bibinfo{person}{Christoph Steinbeck}.}
  \bibinfo{year}{2021}\natexlab{}.
\newblock \showarticletitle{{{COCONUT}} Online: {{Collection}} of {{Open
  Natural Products}} Database}.
\newblock \bibinfo{journal}{\emph{Journal of Cheminformatics}}
  \bibinfo{volume}{13}, \bibinfo{number}{1} (\bibinfo{year}{2021}),
  \bibinfo{pages}{2}.
\newblock
\urldef\tempurl%
\url{https://doi.org/10.1186/s13321-020-00478-9}
\showDOI{\tempurl}


\bibitem[Walker et~al\mbox{.}(2024)]%
        {Walker2024-experimentallymeasured}
\bibfield{author}{\bibinfo{person}{Sara~I. Walker}, \bibinfo{person}{Cole
  Mathis}, \bibinfo{person}{Stuart Marshall}, {and} \bibinfo{person}{Leroy
  Cronin}.} \bibinfo{year}{2024}\natexlab{}.
\newblock \showarticletitle{Experimentally Measured Assembly Indices Are
  Required to Determine the Threshold for Life}.
\newblock \bibinfo{journal}{\emph{Journal of The Royal Society Interface}}
  \bibinfo{volume}{21}, \bibinfo{number}{220} (\bibinfo{year}{2024}),
  \bibinfo{pages}{20240367}.
\newblock
\urldef\tempurl%
\url{https://doi.org/10.1098/rsif.2024.0367}
\showDOI{\tempurl}


\end{thebibliography}

\end{document}